\documentclass[11pt]{article}
\usepackage{geometry}                
\usepackage{graphicx}
\usepackage{amssymb}
\usepackage{amsmath}
\usepackage{mathtext}
\usepackage{float}
\usepackage{afterpage}
\usepackage{wrapfig}
\usepackage[style=nature,backend=bibtex8,maxnames=2,uniquelist=false]{biblatex}
\newcommand{\bi}{\begin{itemize}}
\newcommand{\ei}{\end{itemize}}
\newcommand{\be}{\begin{equation}}
\newcommand{\ee}{\end{equation}}

\newcommand{\bc}{\begin{columns}} 
\newcommand{\ec}{\end{columns}}

\newcommand{\pr}{\text{Pr}}
\newcommand{\prem}{\text{Pr}^\text{em}}

 \usepackage{tabularx,booktabs}
\usepackage[dvipsnames]{xcolor}
\usepackage{tikz}
\usetikzlibrary {fit}
\usetikzlibrary {shapes.geometric}
\usetikzlibrary{calc}

\definecolor{mlila}{RGB}{113, 46, 103}
\definecolor{mblue}{RGB}{38, 117, 146}
\definecolor{mgreen}{RGB}{95, 177, 42}
\definecolor{myellow}{RGB}{199, 200, 0}
\definecolor{morange}{RGB}{255, 121, 23}
\definecolor{mred}{RGB}{226, 58, 52}
\definecolor{mbrown}{RGB}{177,75,39}
\definecolor{mviolet}{RGB}{125,99,193}
\definecolor{msky}{RGB}{92,166,251}
\definecolor{mlime}{RGB}{92,185,111}
\definecolor{mscarlett}{RGB}{234,80,73}
\definecolor{mlemon}{RGB}{255,253,79}

\begin{document}

\title{Hypergraph Animals} 
\author{\textsc{Michael P.H. Stumpf}\\School of BioSciences, University of Melbourne\\
School of Mathematics and Statistics, University of Melbourne\\
\texttt{mstumpf@unimelb.edu.au}}
\maketitle

\begin{abstract}
Here we introduce simple structures for the analysis of complex hypergraphs, hypergraph animals. These structures are designed to describe the local node neighbourhoods of nodes in hypergraphs. We establish their relationships to lattice animals and network motifs, and we develop their combinatorial properties for sparse and uncorrelated hypergraphs. We  make use of the tight link of hypergraph animals to partition numbers, which opens up a vast mathematical framework for the analysis of hypergraph animals. We then study their abundances in random hypergraphs.  Two transferable insights result from this analysis: (i) it establishes the importance of high-cardinality edges in ensembles of random hypergraphs that are inspired by the classical Erd\"os-Reny\'i random graphs; and (ii) there is a  close connection between degree and hyperedge cardinality in random hypergraphs that shapes animal abundances and spectra profoundly. Both findings imply that hypergraph animals can have the potential to affect information flow and processing in complex systems. Our analysis of also suggests  that we need to spend more effort on investigating and developing suitable conditional ensembles of random hypergraphs that can capture real-world structures and their complex dependency structures.  
\end{abstract}




\section{Introduction}
Complex systems are more than the sum of their parts \cite{anderson1972more}. While they cannot be fully comprehended by dissecting them into their constituent components \cite{stumpf2022more}, we can gain valuable, if partial, insights into their global dynamics by characterising key local features. The search and classification of such local features has been an important cornerstone of complex systems science.  Here I consider one candidate feature for complex hypergraphs \cite{battiston2021physics,Vittadello:2023aa}, and which I will refer to as ``hypergraph animals". Below I will outline their relationships to lattice animals and network motifs; I will then discuss some of their combinatorial properties before deriving and analysing their distributions in a class of random hypergraphs. 
\par
Lattice animals \cite{domb1976lattice}  refer to connected groups of  sites in lattices or hypercubes (see Fig. 1A). The number of different configurations for a fixed number of lattice sites, their volume, and circumference of lattice animals have been a topic of longstanding interest in statistical physics, combinatorics, experimental mathematics, but also in recreational mathematics \cite{jensen2000statistics}. The study of lattice animals and their growth has been important in percolation theory to map out percolation transitions on different lattices. 
\par
Network motifs \cite{Milo:2002aa,Stone:2019aa,Mann:2022aa} (Fig. 1B) are a closely related concept in the areas of graph theory and network science.  There are different ways of defining motifs and we follow the definition in  \cite{Alon:2007p9767}: a motif is understood as a small set of nodes connected in a recurring pattern of interactions.  Here, in addition to the combinatorial problems of enumerating motifs or calculating distributions over motif abundances  for different random graph ensembles, the potential functional role that different motifs play in complex dynamical systems has guided much of the research \cite{ingram2006network,Burgio:2021aa,Mann:2023aa}.  A functional pivotally important motif may be rare, but that can only be determined with the correct statistical reference framework to assess statistical significance (or lack thereof) of motif abundances \cite{Thorne:2007is,Karrer:2010aa,Shan:2022aa}. 
\par
 Graphs and networks have come to predominate the analysis of many complex systems\cite{dorogovtsev2022nature}, ranging from physics \cite{cimini2019statistical}, engineering \cite{cui2010complex}, biology \cite{desilva2005complex}, the social sciences \cite{an2022causal}, and the humanities \cite{laidlaw2013colonial}. As our understanding of complex systems and complex networks has increased, so has our appreciation of the potential shortcomings of classical graph descriptions. In many instances interactions and dynamics go beyond pairs \cite{oster1971network,battiston2020networks,vittadello2021model,Vittadello:2023aa,Wegner:2021aa}. Many molecular reactions, for example, involve more than two molecular species; regulation of cellular processes \cite{Wong:2020aa}  and biological information transmission along molecular reaction systems \cite{Mahon:2015hz} rely on the intricate and synergistic stochastic interplay of transcription regulators and their molecular complexes with DNA and co-factors\cite{Ham:2024aa}; in the social sciences dynamics are shaped by the confluence of multiple players that are only poorly understood in terms of pairwise interactions or networks\cite{Muller:2022aa}.  
 \par
The purpose here is to introduce a set of  combinatorial structures that are related to lattice animals and to network motifs, and which for convenience I refer to as {\em hypergraph animals} (see Fig. 1C). I develop their connections to lattice animals and network motifs, and explore their combinatorial properties and their abundances in random uncorrelated hypergraph models \cite{diaz2022hypergraphs}.  Unlike the recent proposition of hypergraph motifs \cite{lee2020hypergraph,lotito2022higher,Wegner:2021aa}, hypergraph animals focus on very local properties: the arrangement of hyperedges incident on a single node. Like for motifs, we have to distinguish between the functional characteristics of a hypergraph animal (e.g. in information transmission  \cite{Mahon:2015hz}, evolutionary graph theory \cite{Altrock:2017aa}, or the regulation of cellular behaviour \cite{Araujo:2023aa}) and the statistical distributions of hypergraph animals in real hypergraphs, or in theoretical hypergraph ensembles. Here I focus on the latter to lay the foundations for a statistical analysis framework for hypergraph animals. 
\par
What all three structures --- lattice animals, network motifs, and hypergraph animals --- have in common is that they focus on local aspects of larger, and potentially more complex structures. There is hope, often with good reason\cite{Alon:2007p9767,May:2010ft,dorogovtsev2022nature}, that understanding local characteristics allows us to understand aspects of the global system. But whether hypergraph animals are such useful local features will depend on their functional properties and their distributions in theoretical as well as empirical hypergraph networks. The analysis of hypergraphs in real-world  applications will be considered in seperate publications.


Hypergraphs  \cite{carletti2020dynamical,dorogovtsev2022nature,battiston2020networks,Arregui-Garcia:2024aa} extend the concept of graphs by allowing edges to contain more than two vertices: edges are sets (or multisets if a node can be part of a hyperedge multiple times) of vertices, $\mathcal{H}=\{h_1,\ldots,h_M\}$ with $h_i=(v_s,v_t,\ldots,v_u)$; we thus have\cite{berge1984hypergraphs,bretto2013hypergraph},
\begin{equation}
\mathcal{J}=(\mathcal{V},\mathcal{H}).
\label{eq:hypergraph} 
\end{equation}
 Here, we refer to $N=|\mathcal{V}|$ as the order of a (hyper)graph, to $M=|\mathcal{E}|$ and $M=|\mathcal{H}|$ as the size of the hypergraph, respectively, and to $l_i=|h_i|$ as the cardinality of hyperedge $i$.  A node, $i\in \mathcal{V}$, that belongs to $k$ hyperedges has degree $k$, and the total number of neighbours is given by the sum over the distinct nodes  that are belong to its $k$ incident hyperedges. For sparse and uncorrelated hypergraphs where hyperedges do not overlap in more that one node this the number of neighbours is given by
 \be
 q_i = \sum_{j|i\in h_j} (l_j-1);
 \ee 
 otherwise the number of neighbours is the number of distinct elements in the multiset $\cup_{j|i\in h_j} h_i$.
  \par
 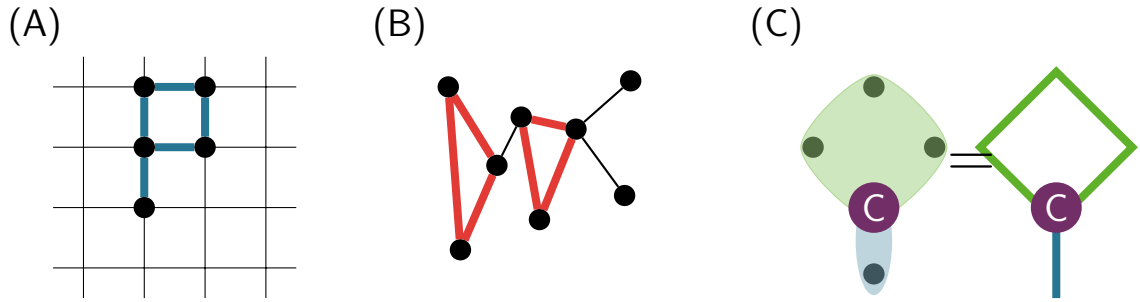
\begin{figure}
 \hskip-9mm
            \begin{tikzpicture}[
dot/.style = {circle, fill, minimum size=#1,
              inner sep=0pt, outer sep=0pt},
dot/.default = 8pt,scale=0.8]  
\draw[step=1cm,black,thin] (-2.5,-2.5) grid (1.5,1.5);
\node[dot] (n1) at (-1,-1) {};
\node[dot] (n2) at (-1,0) {};
\node[dot] (n3) at (-1,1) {};
\node[dot] (n4) at (0,1) {};
\node[dot] (n5) at (0,0) {};
\draw[line width=3pt,mblue] (n1) -- (n2) -- (n3) --(n4) -- (n5) -- (n2);


\node[dot] (g1) at (4,1) {};
\node[dot] (g2) at (6.1,0.3) {};
\node[dot] (g3) at (4.8,-0.3) {};
\node[dot] (g4) at (5.2,0.5) {};
\node[dot] (g5) at (5.5,-1.2) {};
\node[dot] (g6) at (4.2,-1.7) {};
\node[dot] (g7) at (6.9,-0.8) {};
\draw[thick] (g1) -- (g3) --(g4)-- (g5) -- (g2)  --(g4) ;
\draw[thick] (g2) -- (g7); 
\draw[thick] (g1) -- (g6) ; 
\draw[line width=3pt,mgreen] (g4) -- (g5) -- (g3)-- (g2) -- (g5)  -- (g3) --(g4)--(g2) ;

\node[dot] (h1) at (11, -2.1) {};
\node[dot] (h3) at (10, -0) {};
\node[dot] (h4) at (12, -0.) {};
\node[dot] (h5) at (11, 1.) {};
\filldraw[mgreen,fill=mgreen!60,opacity=0.5] plot[smooth cycle] coordinates{(11,-1.2) (12.2,0)  (11,1.2)  (9.8,.0)};
 \fill[mblue,fill=mblue!60,opacity=0.5]  (11,-1.55) circle [x radius=0.3cm, y radius=9mm];
\node[circle,radius=3pt,fill=mlila,inner sep=2pt] (h2) at (11, -1) {\Large \color{white}\sffamily C};
\node[line width=3pt, regular polygon, regular polygon sides=4, minimum size=2cm,mgreen, draw,rotate=45] (Q1) at (14.5,0.) {};
 \draw[line width=3pt, mblue]  (14.5,-1) -- (14.5,-2.5);
 \node[circle,radius=3pt,fill=mlila,inner sep=2pt ] (5A3) at  (14.5,-1)  {\Large \color{white} \sffamily C};
\node[black] (equal) at (12.7,-0.1) {\Huge =};
 \node (A) at  (-2.8,2)  {\Large \sffamily (A)};
 \node (B) at  (3.2,2)  {\Large \sffamily (B)};
 \node (C) at  (9.4,2)  {\Large \sffamily (C)};

\end{tikzpicture}
\caption{\sffamily Structures on lattices, graphs and hypergraphs. (A) {Lattice Animals} are structures formed by connected sites, traditionally on regular lattice. (B) Network motifs are patterns of connected nodes on graphs or networks; their respective abundances and structures reflect structural and potentially functional characteristics of  complex networks. (C) Hypergraph Animals are considered here in an attempt to classify the local properties of nodes (here the central  node ``C" on which the hypergraph animal in incident is indicated in purple) in a hypergraph, where the local neighbourhood is not well captured by the node degree alone. The notation on the right will be used below to keep track of the different potential architectures of Hypergraph Animals. }
\end{figure}

 \par
 \section{Counting Lattice Animal Configurations}
 
A node in a hypergraph with degree $k$ can have $q\ge k$ neighbours; any edge with cardinality $l_i$ contributes $l_i-1$ neighbours. I start by assuming that the hypergraphs are uncorrelated and sufficiently sparse so that we never have the case where the intersection of two vertex sets of  hyperedges  contains more than one element. 
Therefore the number of distinct local configurations in which $q$ neighbours can be contributed by $k$ distinct edges is equal to the so-called partition number, $n(q,k)$, which is the number of ways in which $k$ integers, $0\le k \le q$ can be added up to sum up to $q$. E.g there are $n(4,3) = 3$ ways in which we can sum up three integers to yield 4: $0$ and $4$, $1$ and $3$; and $2$ and $2$ (the order of terms does not matter).  
 \par
 There is no closed form expression for partition numbers, but they obey the recursion relation
 \begin{equation}
 n(q,k) = n(q-1,k-1) + n(q-k,k) 
 \end{equation}
 with
 \begin{equation}
 n(q,k) = 1 \ \   \text{ for } \ \ k =1 \qquad \qquad \text { and } \qquad \qquad n(q,k) = 0 \ \ \text{ for } \ \ n < 0. 
 \end{equation}
In Fig. \ref{fig:animals} we see the possible hypergraph animals for $2\le q \le 5$ in our shorthand notation. This notation is straightforwardly translated into e.g. Young (or Ferrers) diagrams\cite{Schroeder:2006aa,Graham:1994aa}, by adding a row of with $\nu-1$ squares, where $\nu$ is the cardinality of the hyperedge or here the number of sides of the polygons making up the hypergraph animal. The number of rows is the number of hyperedges incident on the anchor node of the hypergraph animal.  For example, we can identify the Young diagram for a given hypergraph animal, such as
\[
\begin{tikzpicture}[
dot/.style = {circle, fill, minimum size=#1,
              inner sep=0pt, outer sep=0pt},
dot/.default = 4pt]  
\node[thick, regular polygon, regular polygon sides=5, minimum size=0.4cm,mlila, draw,rotate=360/10] (P1) at (0,0.23) {};
       \node[thick, regular polygon, regular polygon sides=3,inner sep=0.cm,minimum size=0.5cm,mred, draw,rotate=270] (T1) at (0.14,-0.25) {};
    \foreach \angle in {-60,-30}
 \draw[rotate around={\angle:(0,0)},mblue,thick](0,0) -- (0,-0.5);
 \node[dot] (A) at  (0,0)  {};
 \node at (1,-0) (eq) {$\equiv$};
 \draw[step=0.5cm,black,thin] (1.99,0.5) grid (4.,1);
  \draw[step=0.5cm,black,thin] (1.99,0) grid (3,0.5);
    \draw[step=0.5cm,black,thin] (1.99,-0.5) grid (2.5,0);
        \draw[step=0.5cm,black,thin] (1.99,-1.) grid (2.5,-0.5);
\end{tikzpicture}
\]
Because of this equivalence we can draw on a vast array of tools and body of mathematical knowledge in order to count and analyse the different shapes of hypergraph animals \cite{Alder:1969,Schroeder:2006aa}.
\par
Partition numbers have been a fruitful field of research in combinatorics and number theory. For example, there is an exact solution for the total number of partitions, $n(q)$, in which $q$ can be represented,  but this solution involves infinite series and is hard to represent \cite{Rademacher:1937aa}. The generating function\cite{Lando:2003aa}, $f(x)=\sum n(q) x^q$, by contrast is simple,
\be
f(x) = \prod_{\nu=1}^\infty \frac{1}{(1-x^\nu)} \qquad \qquad \text{ with } |x|<1,
\label{eq:genfunc1}
\ee
and was already known to Euler \cite{Graham:1994aa,Schroeder:2006aa}.

\begin{figure}[t]
            \begin{tikzpicture}[
dot/.style = {circle, fill, minimum size=#1,
              inner sep=0pt, outer sep=0pt},
dot/.default = 2pt,  
 scale=0.75                   ] 
                    \foreach \angle in {90,270}
 \draw[rotate around={\angle:(0,8)},mblue,very thick](0,8) -- (0,7);
 \node[circle,radius=1pt,fill=black] (Am1) at  (0,8)  {};
  \node[very thick, regular polygon, regular polygon sides=3, minimum size=1cm,mred, draw,rotate=180] (T1) at (3,8.6) {};
   \node[circle,radius=1pt,fill=black] (Am1) at  (3,8)  {};
\node[text centered] (t1) at (0,6.75) {$q=2,k=2$};
\foreach \angle in {0,120,240}
 \draw[rotate around={\angle:(0,4)},mblue,very thick](0,4) -- (0,3);
 \node[circle,radius=1pt,fill=black] (Am1) at  (0,4)  {};
\node[text centered] (t1) at (3,6.75) {$q=2,k=1$};
\foreach \angle in {0}
 \draw[rotate around={\angle:(3,4)},mblue,very thick](3,4) -- (3,3);
   \node[very thick, regular polygon, regular polygon sides=3, minimum size=1cm,mred, draw,rotate=180] (T1) at (3,4.6) {};
   \node[circle,radius=1pt,fill=black] (Am1) at  (3,4)  {};
\node[text centered] (t1) at (0,2.75) {$q=3,k=3$};
\node[very thick, regular polygon, regular polygon sides=4, minimum size=1.2cm,mgreen, draw,rotate=45] (Q1) at (6,4.6) {};
\node[circle,radius=1pt,fill=black] (A3) at  (6,4)  {};
\node[text centered] (t1) at (3,2.75) {$q=3,k=2$};
\node[text centered] (t1) at (6,2.75) {$q=3
,k=1$};
\node[circle,radius=1pt,fill=black] (A) at  (0,0)  {};
\node (B) at  (1.,0.)  {};
\node(C) at  (0,1)  {};
\node (D) at  (-1,0.)  {};
\node (E) at  (0,-1)  {};
\draw[very thick,mblue] (A) -- (B);
\draw[very thick,mblue] (A) -- (C);
\draw[very thick,mblue] (A) -- (D);
\draw[very thick,mblue] (A) -- (E);
\node[circle,radius=1pt,fill=black] (A1) at  (3,0)  {};
\node (B1) at  (3+1.,0.)  {};
\node(C1) at  (3+0,1)  {};
\node (D1) at  (3-1,0.)  {};
\node (E1) at  (3+0,-1)  {};
\draw[very thick,mblue] (A1) -- (D1);
\draw[very thick,mblue] (A1) -- (E1);
  \node[very thick, regular polygon, regular polygon sides=3, minimum size=1cm,mred, draw,rotate=20] (T1) at (3.4,0.4) {};
\node[circle,radius=1pt,fill=black] (A1) at  (3,0)  {};
\node[circle,radius=1pt,fill=black] (A2) at  (6,0)  {};
  \node[very thick, regular polygon, regular polygon sides=3, minimum size=1cm,mred, draw,rotate=20] (T1) at (6.4,0.4) {};
\node[very thick, regular polygon, regular polygon sides=3, minimum size=1cm,mred, draw,rotate=-40] (T1) at (5.6,-0.4) {};
\node[circle,radius=1pt,fill=black] (A2) at  (6,0)  {};
\node[circle,radius=1pt,fill=black] (A3) at  (9,0)  {};
\node (E3) at  (9+0,-1)  {};
\draw[very thick,mblue] (A3) -- (E3);
\node[very thick, regular polygon, regular polygon sides=4, minimum size=1.2cm,mgreen, draw,rotate=45] (Q1) at (9,0.6) {};
\node[circle,radius=1pt,fill=black] (A3) at  (9,0)  {};
\node[circle,radius=1pt,fill=black] (A4) at  (12,0)  {};
\node[very thick, regular polygon, regular polygon sides=5, minimum size=1.3cm,mlila, draw,rotate=360/10] (P1) at (12,0.8) {};
\node[circle,radius=1pt,fill=black] (A4) at  (12,0)  {};
\node[circle,radius=1pt,fill=black] (5A1) at  (0,-4)  {};
\node[text centered] (t1) at (0,-1.35) {$q=4,k=4$};
\node[text centered] (t1) at (3,-1.35) {$q=4,k=3$};
\node[text centered] (t1) at (6,-1.35) {$q=4,k=2$};
\node[text centered] (t1) at (9,-1.35) {$q=4,k=2$};
\node[text centered] (t1) at (12,-1.35) {$q=4,k=1$};
\foreach \angle in {0,360/5,2*360/5,3*360/5,4*360/5}
 \draw[rotate around={\angle:(0,-4)},mblue,very thick](0,-4) -- (0,-5);
 \node[circle,radius=1pt,fill=black] (5A1) at  (0,-4)  {};
\node[circle,radius=1pt,fill=black] (5A2) at  (3,-4)  {};
\foreach \angle in {0,360/5,4*360/5}
 \draw[rotate around={\angle:(3,-4)},mblue,very thick](3,-4) -- (3,-5);
  \node[very thick, regular polygon, regular polygon sides=3, minimum size=1cm,mred, draw,rotate=180] (T1) at (3,-3.4) {};
 \node[circle,radius=1pt,fill=black] (5A2) at  (3,-4)  {};
 \node[circle,radius=1pt,fill=black] (5A3) at  (6,-4)  {};
\foreach \angle in {0}
 \draw[rotate around={\angle:(6,-4)},mblue,very thick](6,-4) -- (6,-5);
 \node[very thick, regular polygon, regular polygon sides=3, minimum size=1cm,mred, draw,rotate=-120] (T1) at (6.5,-3.7) {};
  \node[very thick, regular polygon, regular polygon sides=3, minimum size=1cm,mred, draw,rotate=-120] (T1) at (5.5,-3.7) {};
 \node[circle,radius=1pt,fill=black] (5A3) at  (6,-4)  {};
\node[circle,radius=1pt,fill=black] (5A4) at  (12,-4)  {};
\foreach \angle in {-40,40}
 \draw[rotate around={\angle:(9,-4)},mblue,very thick](9,-4) -- (9,-5);
    \node[very thick, regular polygon, regular polygon sides=4, minimum size=1.2cm,mgreen, draw,rotate=45] (Q1) at (9,-3.4) {};
 \node[circle,radius=1pt,fill=black] (5A3) at  (9,-4)  {};
\node[circle,radius=1pt,fill=black] (5A4) at  (12,-4)  {};
  \node[very thick, regular polygon, regular polygon sides=3, minimum size=1cm,mred, draw,rotate=0] (T1) at (12,-4.6) {};
   \node[very thick, regular polygon, regular polygon sides=4, minimum size=1.2cm,mgreen, draw,rotate=45] (Q1) at (12,-3.4) {};
\node[circle,radius=1pt,fill=black] (5A4) at  (12,-4)  {};
\node[circle,radius=1pt,fill=black] (A4) at  (15,-4)  {};
\foreach \angle in {0}
 \draw[rotate around={\angle:(15,-4)},mblue,very thick](15,-4) -- (15,-5);
\node[very thick, regular polygon, regular polygon sides=5, minimum size=1.3cm,mlila, draw,rotate=360/10] (P1) at (15,-3.2) {};
\node[circle,radius=1pt,fill=black] (A4) at  (15,0-4)  {};
\node[very thick, regular polygon, regular polygon sides=6, minimum size=1.3cm,morange, draw,rotate=30] (P1) at (18,-3.2) {};
\node[circle,radius=1pt,fill=black] (A4) at  (18,-4)  {};
\node[text centered] (t1) at (0,-5.35) {$q=5,k=5$};
\node[text centered] (t1) at (3,-5.35) {$q=5,k=4$};
\node[text centered] (t1) at (6,-5.35) {$q=5,k=3$};
\node[text centered] (t1) at (9,-5.35) {$q=5,k=3$};
\node[text centered] (t1) at (12,-5.35) {$q=5,k=2$};
\node[text centered] (t1) at (15,-5.35) {$q=5,k=2$};
\node[text centered] (t1) at (18,-5.35) {$q=5,k=1$};
\end{tikzpicture}
\caption{\sffamily Distinct hypergraph animals for $2\le q\le 5$. Here $q$ refers to the total number of distinct neighbours and $k$ the number of edges incident upon the central anchor node of the hypergraph animal.}
\label{fig:animals}
\end{figure}

\subsection{Restricted Hypergraph Animals}
The combinatorial properties of partition numbers carry through to hypergraph animals. We can, for example, calculate the number of animals that have no edges of cardinality 2, that is nodes for which  every incident edge is a true hyperedge ($l_1>2$); below I will show that these are often dominating the set of observed hypergraph animals in random hypergraphs. In this case the generating function becomes,
\begin{equation}
f_1(x) =\prod_{\nu=2}^\infty \frac{1}{(1-x^\nu)} \qquad \qquad \text{ with } |x|<1.
\label{eq:genfunc2}
\end{equation}
The proof of this is analogous to the standard proof of relationship \eqref{eq:genfunc1} in e.g. \cite{Alder:1969,Schroeder:2006aa} and proceeds by expanding the right-hand side of Eqn. \eqref{eq:genfunc2} as a power series,
\begin{align*}
\prod_{\nu=2}^\infty \frac{1}{(1-x^\nu)} \qquad =\qquad&  (1+x^2+x^{2\cdot2}+x^{2\cdot3}+\ldots)\times\\
& (1+x^3+x^{3\cdot2}+x^{3\cdot3}+\ldots)\times\\
& (1+x^4+x^{4\cdot2}+x^{4\cdot3}+\ldots)\times\\
& (1+x^5+x^{5\cdot2}+x^{5\cdot3}+\ldots)\times\\
& (1+x^6+x^{6\cdot2}+x^{6\cdot3}+\ldots)\times\\
 &\ldots.
\end{align*}
Collecting terms of equal power, for example for $q=5$, we have the following ways of obtaining $x^5$,
\[
1\times 1\times 1\times x^5\qquad \text{ and }\qquad  x^2\times x^3\times 1.
\]
Thus the only two conformations that result in 5 neighbours only involving hyperedges with $l_i>2$ are \\[2mm]
\begin{tikzpicture}[
dot/.style = {circle, fill, minimum size=#1,
              inner sep=0pt, outer sep=0pt},
dot/.default = 4pt]  
  \node[very thick, regular polygon, regular polygon sides=3, minimum size=0.8cm,mred, draw,rotate=0] (T1) at (3,-0.4) {};
   \node[very thick, regular polygon, regular polygon sides=4, minimum size=0.8cm,mgreen, draw,rotate=45] (Q1) at (3,0.4) {};
\node (text) at (1.5,0.1) {and};
       \node[very thick, regular polygon, regular polygon sides=6, minimum size=0.9cm,morange, draw,rotate=30] (P1) at (0,0.1) {};

\end{tikzpicture}
 
Similarly, for $q=6$ we have four potential configurations,
\[
1\times 1\times 1\times1 \times  x^6\text{,}\qquad  x^2\times 1 \times x^4\times 1\text{,}\qquad  1\times x^{3\cdot2}\times 1\qquad\text{ and } \qquad  x^{2\cdot3}\times 1,
\]
or, in our graphical notation,\\[2mm]
\begin{tikzpicture}[
dot/.style = {circle, fill, minimum size=#1,
              inner sep=0pt, outer sep=0pt},
dot/.default = 4pt]  
  \node[very thick, regular polygon, regular polygon sides=3, minimum size=0.8cm,mred, draw,rotate=0] (T1) at (3,-0.4) {};
   \node[very thick, regular polygon, regular polygon sides=5, minimum size=0.8cm,mlila, draw,rotate=360/10] (Q1) at (3,0.4) {};
       \node[very thick, regular polygon, regular polygon sides=7, minimum size=0.9cm,gray, draw,rotate=360/14] (P1) at (0,0.1) {};
   \node[very thick, regular polygon, regular polygon sides=4, minimum size=0.8cm,mgreen, draw,rotate=45] (Q1) at (5.6,0.1) {};
   \node[very thick, regular polygon, regular polygon sides=4, minimum size=0.8cm,mgreen, draw,rotate=45] (Q1) at (6.4,0.1) {};
\node (text) at (7.5,0.1) {and};
  \node[very thick, regular polygon, regular polygon sides=3, minimum size=0.8cm,mred, draw,rotate=0] (T1) at (9,-0.4) {};
  \node[very thick, regular polygon, regular polygon sides=3, minimum size=0.8cm,mred, draw,rotate=120] (T1) at (8.65,0.2) {};
  \node[very thick, regular polygon, regular polygon sides=3, minimum size=0.8cm,mred, draw,rotate=-120] (T1) at (9.35,0.2) {};
\node (point) at (9.9,0.1) {.};

\end{tikzpicture}
\\[2mm]
Eqns. \eqref{eq:genfunc1} and \eqref{eq:genfunc2} can be generalised to calculate the number of abundances of any restriction on hypegraph animals. For example, if we want to consider animals which only consist of hyperedges of cardinalities $l_i \in \mathcal{L}=\{\lambda_1,\lambda_2,\lambda_3,\ldots,\lambda_s\}$ we have for the generating function 
\begin{equation}
f_2(x) = \prod_{\nu \in \mathcal{L}} \frac{1}{(1-x^\nu)}
\end{equation} 
The generating functions of this type allow us to calculate the number of different hypergraph animal configurations.
 To calculate their abundances in uncorrelated hypergraphs we need to define suitable hypergraph ensembles, which I will turn to in the next section. 

\section{Hypergraph Animals in  Random Hypergraphs}
\subsection{Random Hyperghraph Ensembles}
\afterpage{
\begin{figure}[H]
\vskip-3mm
\includegraphics[width=1\textwidth]{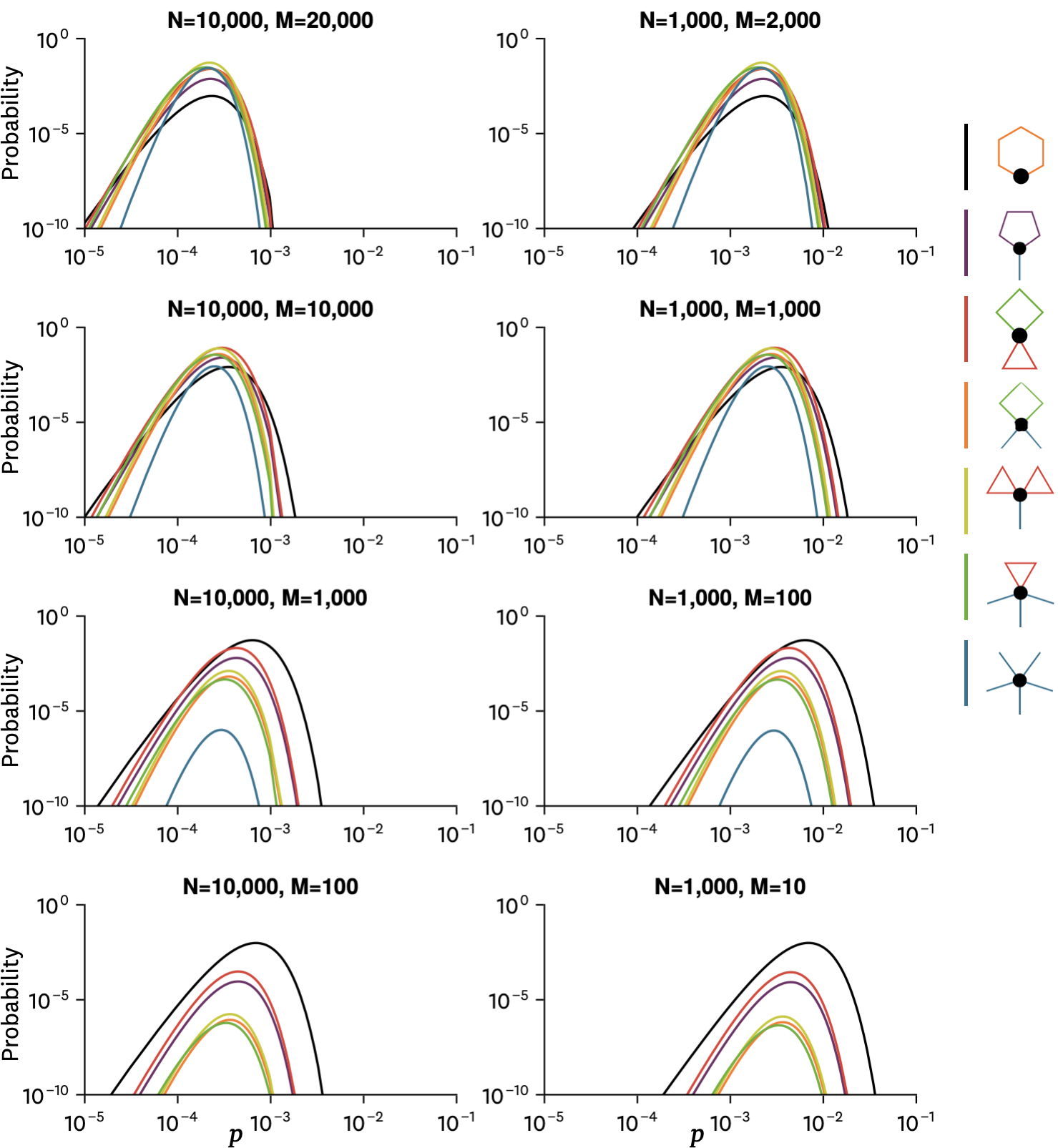}
\label{fig:animaldists}
\caption{\sffamily  Hypergraph-animal probabilities for $q=5$ vs. the probability, $p$, of a node $i$ belonging to any given hyper edge $j$ (see Eqns. \eqref{eq:Poideg} and \eqref{eq:Poicard}). The left hand side shows this for $N=10^6$, and the right hand side for $N=10^3$. High-degree hypergraph animals tend to be  rare in random hypergraphs.  
}
\end{figure}
}

We introduce an ensemble of random hypergraphs that follows as closely as possible the example of  the classical or Erdös--Renyi random graphs \cite{Bollobas:2001aa} and which was recently discussed by Barthelemy \cite{Barthelemy:2022aa}. I consider hypergraphs of order $N$ (the number of nodes) and size, $M$ (the number of hyperedges). For $i=1,\ldots,N$ each edge $j$ (with $j=1,\ldots,M$) contains node $i$ with probability $p$. Then the degree distribution, that is the probability that a node is part of  $k$ hyperedges, is 
\begin{equation}
\pr(k) = \binom{M}{k}p^k(1-p)^{M-k}.
\label{eq:RHGdeg}
\end{equation} 
Similarly the distribution over edge cardinalities, that is the probability that a hyperedge contains $l$ distinct vertices (I do not consider cases where hyperedges can be multisets) is given by
\be
 \pr(l) = \binom{N}{l}p^l(1-p)^{N-l}.
 \label{eq:RHGcard}
 \ee
 \par
With Eqns. \eqref{eq:RHGdeg} and \eqref{eq:RHGcard} we can immediately calculate the average degree and cardinality of the hypergraph given $N, M$ and $p$,
 \be
 \langle k\rangle = pM
 \ee
 and 
 \be
 \langle l\rangle = pN,
 \label{eq:avgCard}
 \ee
respectively.  The definitions, \eqref{eq:RHGdeg} and \eqref{eq:RHGcard} of  \cite{Barthelemy:2022aa}, allow unconnected nodes ($k=0$),  hyperedges with only one incident node ($l=1$), as well as empty hyper-edges $(l=0)$. We can choose to rescale these distributions \cite{stumpf2010incomplete} to focus only on nodes with at least one incident edge and edges with at least 2 nodes; the corresponding relationships are shown in appendix A.
\par

 It is sometimes preferable to write $M=\gamma N$, with $\gamma\in \mathbb{R}_0^+$. With this we can, in the limit $N\longrightarrow \infty$, make the transition from binomial to Poisson distributions over degrees and cardinalities,
\begin{equation}
\pr(k) = \frac{(\gamma\lambda)^k \exp(-\gamma\lambda)}{k!}\label{eq:Poideg}
\end{equation} 
and
\begin{equation}
\tilde{\pr}(l) = \frac{(\lambda)^l \exp(-\lambda)}{l!}\label{eq:Poicard},
\end{equation} 
where $\lambda=pN$.

\par
From equations \eqref{eq:RHGcard}-\eqref{eq:Poicard} it follows that the average cardinality is proportional (with proportionality constance $\gamma$) to the average degree;  most nodes will have a degree close to the average degree of the hypergraph; and most hyperedges will have cardinality close to the average cardinality.

\subsection{Hypergraph Animal Abundances in Random Hypergraphs}
In an uncorrelated hypergraph the numbers of  different hypergraph animals can be calculated  following arguments employed e.g. in \cite{stumpf2010incomplete} for random graphs: we need to take the probability distributions over node degrees and edge cardinalities into account, and then consider the combinatorics of assigning $q$ neighbours to $k$ edges.  The dependency between (hyper-)node degrees and cardinality makes this a somewhat richer behaviour than might be expected  from the analysis of random graphs, where  these two are unlinked. As $p$ increases so does $\langle k\rangle$ but also $\langle l\rangle$.  Thus in the simple ensemble of random hypergraphs described above the probability of a hyperedge having cardinality, $l=2$, initially increases and then decreases. That in turn also means that very quickly the fraction of hypergraph animals that are composed solely of cardinality 2 hyperedges decreases: random hypergraphs become dominated by hyperedges with cardinality close to the average cardinality \eqref{eq:avgCard}.
\par
With the probabilities for degrees and cardinalities given we can calculate the probability of any configuration where a node with degree $k$ has $q$ neighbours across its $k$ incident hyperedges with cardinalities, $l_i$, subject to the constraint that $\sum_{i=1}^k (l_i-1) = q$. Assuming that the hypergraph is uncorrelated and that the overlap between hyperedges is never more than one node we obtain,
\be
\pr(k,\{l_i\}) = \binom{q}{l_1-1;l_2-1;\ldots;l_k-1} \pr(k) \tilde{\pr}(l_1)\tilde{\pr}(l_2)\ldots \tilde{\pr}(l_k).
\label{eq:animalprob}
\ee
We can substitute Eqns. \eqref{eq:Poideg} and \eqref{eq:Poicard} (or, alternatively but less conveniently,  expressions \eqref{eq:RHGdeg} and \eqref{eq:RHGcard}) into this expression to determine the probability of any hypergraph animal,
\be
 \pr(k,\{l_i\}) = \binom{q}{l_1-1;l_2-1;\ldots;l_k-1}  \frac{(\gamma\lambda)^k e^{-\gamma\lambda}}{k!}\frac{\lambda^{l_1} e^{-\lambda}}{l_1!}\frac{\lambda^{l_2} e^{-\lambda}}{l_2!}\ldots\frac{\lambda^{l_k} e^{-\lambda}}{l_k!}.
 \label{eq:randprobs}
\ee
We can rewrite this equation in the following form
\be
 \pr(k,\{l_i\})  =   \binom{q}{l_1-1;l_2-1;\ldots;l_k-1} \frac{ \gamma^k\lambda^{q+k} e^{-\lambda(k+\gamma)}}{k!\prod_{i=1}^k l_i!}
\ee
Because the probability depends only on the hypergraph animal characteristics, $k$ and $\{l_i\}$ and on $\lambda = p N$ and $\gamma =M/N$, we will, for different choices of  $N$, say $N_1$ and $N_2$, but fixed $\gamma$ observe the same form for $\pr(k,\{l_i\}) $ with a simple rescaling of $p_2 = N_1 p_1/N_2$. Plotting the probability distributions against $\lambda=p*N$ thus yields a universal curve (for sufficiently large $N$).
\par
For the ensemble definition of random hypergraphs in Eqns. \eqref{eq:animalprob} and  \eqref{eq:randprobs},  it is possible (in fact generally inevitable) to have hyperedges with no or one incident node, and nodes that are not part of any hyperedge. To compare against empirical or simulated hypergraphs we therefore need to normalise the probabilities to make sure that (i) every node has at least degree 1, $k\ge 1$, and every hyperedge has at least two incident nodes, $l\ge 2$. With this we obtain the adjusted probabilities
\be
\prem(k,\{l_i\}) = \frac{\pr(k,\{l_i\}) }{(1-\pr(k=0))\prod_i (1-\tilde{\pr}(l_i<2))},
\ee
which can, with some caution, be used for empirical analyses of hypergaph animal abundances in real hypergraphs. These are further discussed in appendix A. Simulation of hypergraphs is outlined in Appendix B.
\par 
\begin{wrapfigure}{l}{0.6\textwidth}
\vskip-1mm
\includegraphics[width=3.5in]{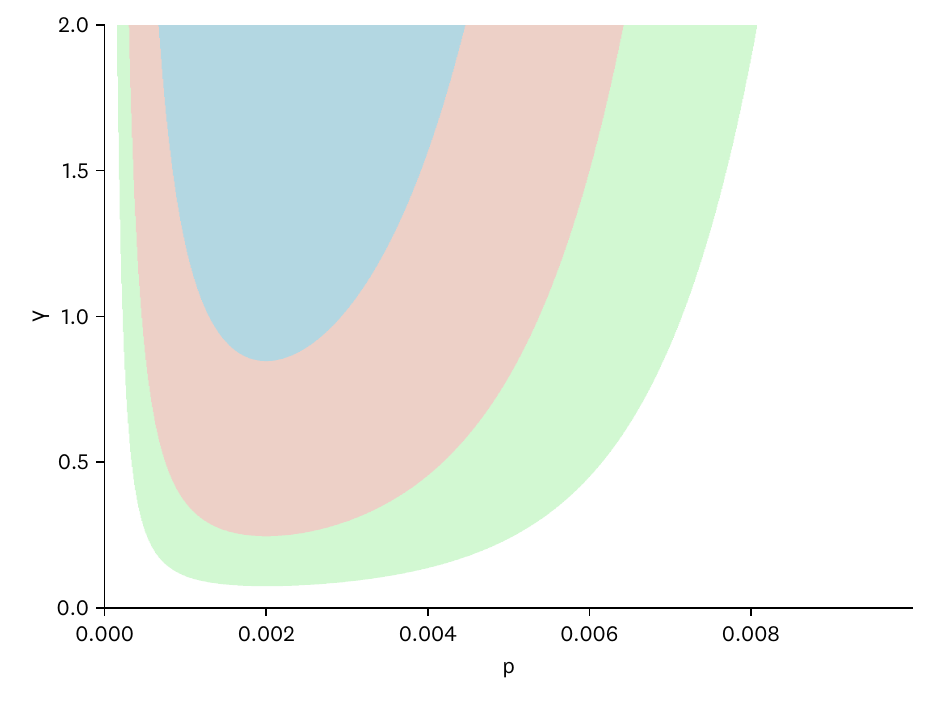}
\vskip-2mm
\caption{\sffamily  The light green area  is the regime in $p\!\! -\!
\! \gamma$ space where  {\rmfamily $\pr(2,\{3,2\}) >\pr(1,\{5\})$}; the pink area indicates where {\rmfamily $\pr(2,\{4,1\}) >\pr(1,\{5\})$}; the light blue area corresponds to the regime where \rmfamily $\pr(5,\{1,1,1,1,1\}) >\pr(1,\{5\})$. }

\end{wrapfigure}

The distributions for animals for which $q=5$ are shown in Fig. 3.  The first noticeable feature is, that as argued above, the shape of the distribution remains identical for the same $\gamma$ and $\lambda$, but shifted for $p$. For moderately sparse, $\gamma = 0.1$, and very sparse , $\gamma =0.01$, hypergraphs, the probability tends to decrease with degree, $k$, of the central node.  For animals with the same $k$, those with the larger minimum hyper edge cardinality have a slightly higher probability over the $p$ ranges considered here.  As $p$ increases the number of neighbours $q$ and the cardinality of hyperedges increases, which, for fixed $q$ leads to an initial increase in the hyper animal probability before it decreases again. 
\par
The structure of Eqn. \eqref{eq:randprobs} helps to explain this: for each additional (hyper)edge incident on a node we gain a Poisson term, ${\lambda^{l_{k+1}} e^{-\lambda}}/{l_{k+1}!}<1.0$, which explains why, generally, higher degree results in a lower hypergraph animal probability for sparse hypergraphs. Similarly, two hypergraph animals with the same degree $k$ but different cardinality sets $\{l_i\}$ and $\{l_j\}$ will typically differ in two (or, especially for large $k$, more) Poisson terms, e.g.
\be
\frac{\lambda^{l_s} e^{-\lambda}}{l_s!}\ldots \frac{\lambda^{l_t} e^{-\lambda}}{l_t!} \qquad \text{ and } \qquad  \frac{\lambda^{l_s-1} e^{-\lambda}}{(l_s-1)!}\ldots \frac{\lambda^{l_t+1} e^{-\lambda}}{(l_t+1)!}; 
\ee
so their products will be of the same order.
\par
For denser hypergraphs, $\gamma \gtrsim 1$, (above the percolation threshold for this type of random hypergraph) the situation changes profoundly, as the probability of low degree nodes decreases. Here  the probability of  the animals is no longer a simple function of degree, but shaped by the interplay between degrees and cardinalities.
\par
As we see in the top-panel of  Fig. 3 there are some areas, where the probability can be higher for $k=2$ than for $k=1$. We can use Eqn. \eqref{eq:randprobs} to map these areas where $\pr(k,\{l_i\}) >\pr(k-1,\{l_j\})$, e.g. 
\be
\pr(2,\{3,2\}) >\pr(1,\{5\}) 
\ee
is fulfilled for fixed $\lambda$ when 
\be 
\gamma >\frac{e^\lambda}{5\lambda}.
\ee
Because $\lambda=Np$ we can determine $\pr(k,\{l_i\}) >\pr(k-1,\{l_j\}) $ as a function of $\gamma$ and $p$.  In Fig. 4  we  use $\pr(2,\{3,2\}) >\pr(1,\{5\})$, $\pr(2,\{4,1\}) >\pr(1,\{5\})$, and   $\pr(2,\{4,1\}) >\pr(5,\{1,1,1,1,1\})$ as illustrations, and indicate  the area where the inequalities hold in light green, pink, and light blue, respectively, which tends to be for larger values of $\gamma$.  For the random hypergraph ensemble considered here we find that generally the frequency of a hypergraph animal decreases with degree $k$, as us seen in Figure 3. The link between node degree and hyperedge cardinality implied by Eqns.  \eqref{eq:Poideg} and \eqref{eq:Poicard}, is a potential shortcoming of this definition of a hypergraph ensemble. In practical applications we should consider configuration hypergraphs \cite{Chodrow_configuration,arafat2020construction}, which can be conditioned on available data.

\section{Conclusion}
Hypergraphs are more complex and flexible than conventional networks or lattices. Hypergraph animals are, in principle, simpler than lattice animals and  network motifs, and they are clearly simpler than recent generalisations of motifs to hypergraphs \cite{lee2020hypergraph,lotito2022higher}.  Nevertheless they hold considerable appeal and promise for further analysis: their combinatorial properties and their probabilistic characteristics distributions in random networks deserve further consideration \cite{Yoon:2011aa}. Here the focus was on uncorrelated and relatively sparse hypergraphs, but  this will cover many important applications \cite{christensen2009reconstruction,battiston2021physics}. In systems biology applications, for example, we can robustly justify the sparsity assumption given available empirical data; and extensive simulations over the range of $\gamma$ values considered here (which are broadly in line with what we observe for many biological networks) the assumptions appear to be justified (code provided on the accompanying GitHub site enables further exploration). The regime where the assumptions made here are no longer valid will be of considerable interest in its own right \cite{Karrer:2010aa,stumpf2010incomplete,Mann:2022aa}, and for compound (or overlapping) hypergraph animals the combinatorial considerations in Section 2 provide a useful starting point. 
\par
One promising application of hypergraph animals is in the statistical analysis of real-world hypergraphs. For a meaningful analysis we need expressions for expected distributions of the different hypergraph animals; and arguably we need ensembles for more general random hypergraphs.  Without the {\em a priori} establishment of a  statistical framework for hypergraph animal analysis, it will be difficult to make meaningful statistical assessments of their association with functional properties of complex hypergraphs. Just as network motifs are only indicative but not conclusive about function and dynamics, so do we have to expect limitations as to how much functional information can be gleaned from hyper graph animal abundances or spectra. They would, however, offer a natural way to define, analyse, and  even infer complex multivariate dependency structures e.g.using suitable information measures\cite{chan2017gene,Chen:2023aa,Shan:2022aa}  from single cell gene expression data.  One of the challenges here will be to enumerate all hypergraph animals efficiently and meaningfully in large networks. This can be challenging if we want to ensure that distinct structures do not overlap (in terms of shared edges).  
\par
The network research community has developed a lot of intuition about networks and their characteristics by combining  rigorous mathematical analysis of random and generalised random graphs, empirical analysis of real-world networks (even though it now appears that many of these will, in fact, be better described by hypergraphs), and the development of error and sampling models for network data. Better understanding relies on all three lines of inquiry supporting one another, and us in our attempts to make sense of complex data, networks, and hypergraphs.  It will therefore be important to progress the development of theoretical and statistical tools alongside, and in support of, the analysis of empirical hypergraph data. Analysis of network data in the absence of suitable statistical tools has given rise to some (not just by hindsight) non-sensical analyses; hypergraphs  are more complicated and therefore more caution and care need to be invested up front, including into the development of powerful null models for random hypergraphs \cite{Chodrow_configuration,Ghoshal:2009aa,Darling:2005aa,Feng:2023aa}. This is especially true for properties and structures derived from hypergraphs, such as the hypergraph animals considered here. 
\par

\section*{Acknowledgements}
I am grateful for the support and helpful comments from the Theoretical Systems Biology group at the University of Melbourne. The work has benefitted greatly from discussions with L\'eo Diaz, Sean Vittadello, and Joshua Forrest. I gratefully acknowledge the support from the Australian Research Council through an ARC Laureate Fellowship (FL220100005). 

\section*{Data Availability}
A Jupyter notebook containing all Julia \cite{Bezanson:2017gd,Roesch:2023aa} code to produce the numericall results in the paper is available at \url{https://github.com/MichaelPHStumpf/HypergraphAnimals}.

\appendix
\section{Adjusted degree and cardinality distributions for empirical random hypergraph ensembles}
Eqns. \eqref{eq:RHGdeg} and \eqref{eq:RHGcard} allow for degree zero  nodes and cardinality zero and one edges, respectively. It makes sense, especially in the context of the analysis of real hypergraph data, to develop hypergraph ensembles where each hyperedge has cardinality $l\ge 2$, and where each node has degree $k\ge 1$. We can rescale the degree and cardinality distributions to develop the adjusted distributions and we obtain, 
\be
\tilde{\pr}(\tilde{k})  = \frac{1}{1-(1-p)^M} \binom{M}{\tilde{k}} p^{\tilde{k}}(1-p)^{M-\tilde{k}}
\label{eq:RHGdegc}
\ee 
and
\be
\tilde{\pr}(\tilde{l})  = \frac{1}{1-(1-p)^N-p(1-p)^{N-1}} \binom{N}{\tilde{l}} p^{\tilde{l}}(1-p)^{N-\tilde{l}}
\label{eq:RHGcardc}
\ee 
The simple scaling means that the qualitative behaviour of equations  \eqref{eq:RHGdedc} and \eqref{eq:RHGcardc} is identical to their counterparts,  Eqns. \eqref{eq:RHGdeg} and \eqref{eq:RHGcard}.  
 \afterpage{
\begin{figure}[H]
\vskip-3mm
\includegraphics[width=1\textwidth]{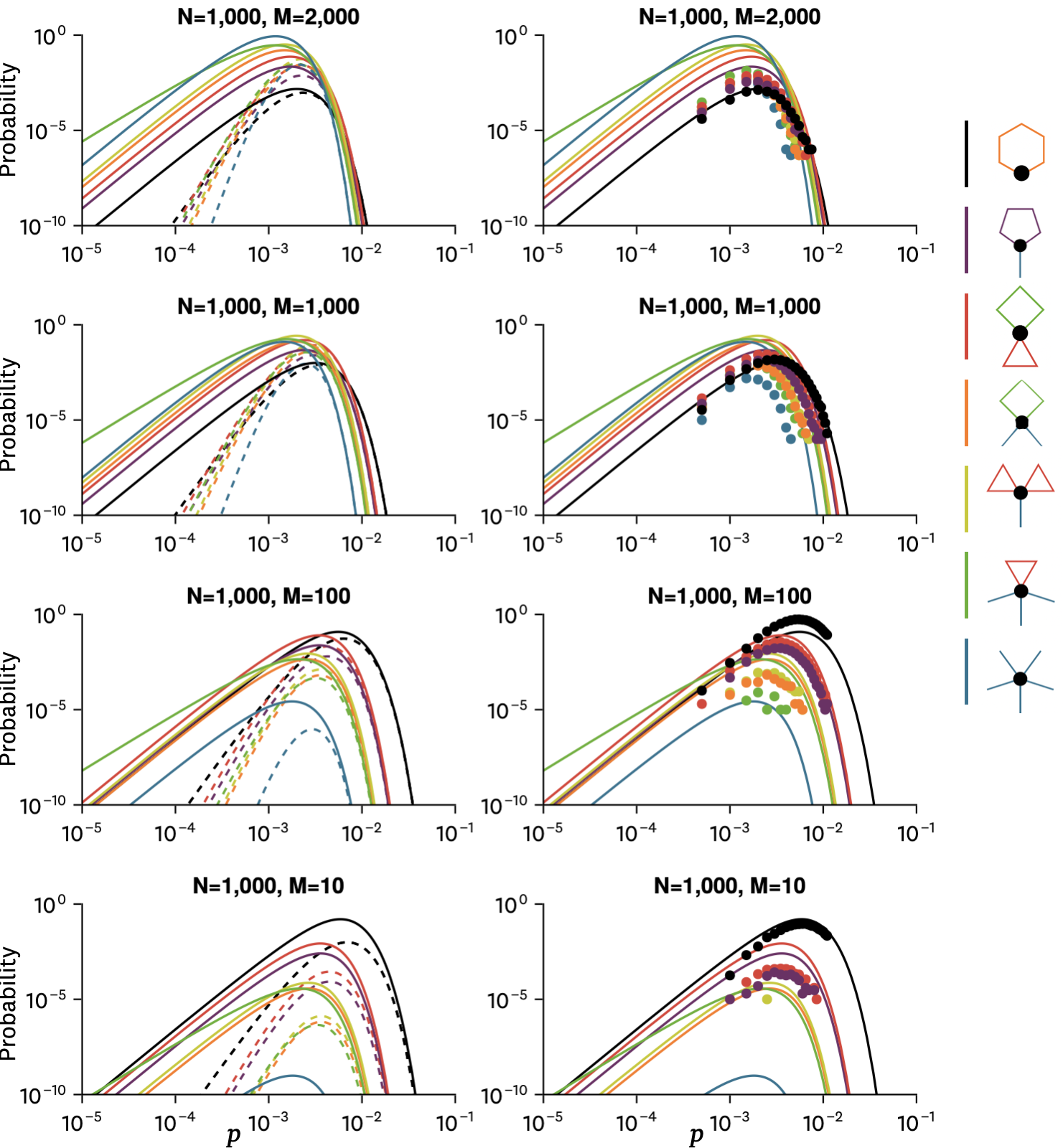}
\label{fig:app}
\caption{\sffamily  Left: Theoretical hypergraph-animal probabilities (dashed lines) and adjusted probabilities (solid lines)  for $q=5$ for different $M/N$ rations. Right: adjusted probabilities (solid lines) and empirical probabilities from simulations of random hypergraphs (dots) . }
\end{figure}
}
\par
In figure \ref{fig:app} I show the agreement between Eqns.  \eqref{eq:RHGdeg} and \eqref{eq:RHGdegc} on the left and between \eqref{eq:RHGdegc} and simulated data  on the right.  The theoretical and adjusted distributions agree for large $p$ but deviate considerably for smaller $p$ . As $\gamma$, the ratio of $M/N$, decreases the adjusted probability distributions deviate more and more from their  theoretical counterparts. 
\par
On the righthand side of the figure we compare the adjusted probabilities to empirical distributions obtained from simulating random hypergraphs. Qualitatively the predictions and simulation results agree. Quantitatively, the agreement is best for moderately sparse graphs: the quantitative discrepancies observed on the righthand side of Figure \ref{fig:app}  arise from the assumption that the hypergraphs are uncorrelated. This can be a problematic assumption for classical graphs, but is even less reliable for the type of random hypergraph ensemble \cite{Barthelemy:2022aa} considered here. In closing we note, that this may cause problems for testing the statistical significance of the abundance of a given hypergraph animal against a suitable null model. In such cases empirical null models should be chosen over theoretical null models.   

\section{Simulating random hypergraph ensembles}
The ensemble as defined by Barthemeley  \cite{Barthelemy:2022aa} is simulated in the Julia programming language as follows:
\begin{enumerate}
\item Choose $N$,$M$, and $p$ the numbers of nodes and hyperedges, and the probability for a node to belong to a hyperedge.
\item Randomly sample the $M$ hyperedge cardinalities $l_i$  from Eqn. \eqref{eq:RHGcard}.
\item For each hyperedge $i$ sample $l_i$ nodes randomly without replacement.
\end{enumerate}
The hypergraph is represented as a set of sets, one for each hyperedge. The code is provided in the Jupyter notebook on GitHub \url{https://github.com/MichaelPHStumpf/HypergraphAnimals}.

\printbibliography{}

 \end{document}